# Structural and magnetic properties of an InGaAs/Fe$_3$Si superlattice in cylindrical geometry

Ch. Deneke,[1,*] J. Schumann,[1] R. Engelhard,[1] J. Thomas,[2] C. Müller,[1] M. S. Khatri,[3] A. Malachias,[4] M. Weisser,[5] T. H. Metzger,[4] and O. G. Schmidt[1]

[1]Institute for Integrative Nanosciences, IFW Dresden, Helmholtzstrasse 20, D-01069 Dresden, Germany

[2]Institute for Complex Materials, IFW Dresden, Helmholtzstrasse 20, D-01069 Dresden, Germany

[3]Institute for Metallic Materials, IFW Dresden, Helmholtzstrasse 20, 01069 Dresden, Germany

[4]European Synchrotron Radiation Facility, Boîte Postale 220, F-38043 Grenoble Cedex, France

[5]Chair for Crystallography and Structural Physics, University of Erlangen-Nürnberg, Staudtstr. 3, 91058 Erlangen, Germany

**Abstract**

The structure and the magnetic properties of an InGaAs/Fe$_3$Si superlattice in a cylindrical geometry are investigated by electron microscopy techniques, x-ray diffraction and magnetometry. To form a radial superlattice, a pseudomorphic InGaAs/Fe$_3$Si bilayer has been released from its substrate self-forming into a rolled-up microtube. Oxide-free interfaces as well as areas of crystalline bonding are observed and an overall lattice mismatch between succeeding layers is determined. The cylindrical symmetry of the final radial superlattice shows a significant effect on the magnetization behavior of the rolled-up layers.

---

[*] e-mail: c.deneke@ifw-dresden.de

## 1. Introduction

The invention and realization of planar superlattices have motivated fundamental as well as application-related research over more than 30 years. Among other properties, the great success of this technology relies on high quality interfaces found especially in epitaxial semiconductor heterostructures. Such interfaces in superlattices allow for the utilization of quantum size effects modifying fundamental optical, electronic or magnetic properties of the heterostructures compared to the bulk material.[1] As classical superlattices are realized mainly by planar growth techniques,[1] they are restricted to planar geometries even if other geometries are of fundamental interest.[2,3] More recently, hybrid semiconductor/magnetic radial superlattices (RSL)[4] have been realized by the roll-up of strained layer systems.[5,6] For potential applications of these RSLs in electrical heterojunctions, tunnel contacts or spin injection devices, the detailed structure of the interfaces as well as the cylindrical geometry will play a major role. For semiconductor-based RSLs the possibility of perfect interfaces has been theoretically investigated[7] and studied previously by transmission electron microscopy (TEM) in detail.[8-12] In these investigations, mostly a disturbed interface region is reported with the exception of the overgrowth on a GaAs (110) cleaved edge, where a dislocated but crystalline bonded interface was found.[12] Also the interfaces of hybrid semiconductor/metal RSLs have been studied in detail[4,13] and normally an oxide-rich interface layer has been observed.

In this work, we investigate the structure and the magnetic properties of an $In_{0.2}Ga_{0.8}As/Fe_3Si$ RSL by complementary experimental techniques. Apart from a thorough study of the interfaces of succeeding periods of the hybrid superlattice, we investigate the lattice mismatch by taking advantage of the cylindrical symmetry of the structure. We find areas of nearly perfect crystalline bonding and an overall lattice mismatch of 2.8% determined by x-ray diffraction. The special cylindrical geometry is also reflected in the magnetic properties of the RSLs changing the symmetry plane of the magnetization axes compared to the planar film.

## 2. Experimental details

An initial $InGaAs/Fe_3Si$ bilayer was fabricated by the combination of III-V molecular beam epitaxy (MBE) and thermal co-evaporation on GaAs (001) substrates. The MBE layers consist of a 10 nm thick AlAs sacrificial layer, followed by 20 nm $In_{0.2}Ga_{0.8}As$. After MBE growth the samples are removed from the ultra-high-vacuum chamber and transferred into a second thermal evaporation chamber. To form an epitaxial bilayer, the MBE-grown III-V surface is thermally deoxidized and a 20 nm $Fe_3Si$ layer is grown pseudomorphically on top of the InGaAs layer by co-evaporation of Fe and Si. Tube formation is initiated *ex-situ* by selective wet chemical removal of the sacrificial AlAs layer with diluted HF. By releasing the inherently strained bilayer from its substrate, it will roll up and form a hybrid $InGaAs/Fe_3Si$ RSL[4] on the sample surface. The rolled-up tube diameter was measured by SEM and an average value of 1200 nm is obtained. To investigate the RSLs structure, tube cross-sections were carefully prepared by focused ion beam (FIB) etching using a Zeiss NVision.[14]

The $InGaAs/Fe_3Si$ RSL cross-section was investigated with energy-filtered transmission electron microscopy (EFTEM) as well as high-angular annual darkfield scanning transmission electron microscopy (HAADF-STEM) in a FEI Tecnai F30 equipped with a Gatan GIF and an energy dispersive x-ray (EDX) detector for chemical analysis. The microscope was operated at an acceleration voltage of 300 kV. Taking advantage of the cylindrical symmetry of the RSL, micro-x-ray diffraction (XRD) measurements on single rolled-up $InGaAs/Fe_3Si$ tubes (located on the same sample used for FIB cross-section preparation) were carried out at the ESRF ID01 beamline at the ESRF, Grenoble. A focused x-ray beam with dimensions of 6 by 7 µm was obtained using Be compound refractive lenses for an energy of 8.8keV (for experimental details see Ref. 15). Magnetic measurements on the planar bilayer and ordered, close-packed ensembles of microtubes were carried out in a Quantum Design physical properties measurement system (PPMS) equipped with a vibrating sample magnetometer (VSM) and magnetic field applied in either parallel or perpendicular direction to the sample (sample size ca. 3x4 mm).

## 3. Results and Discussion

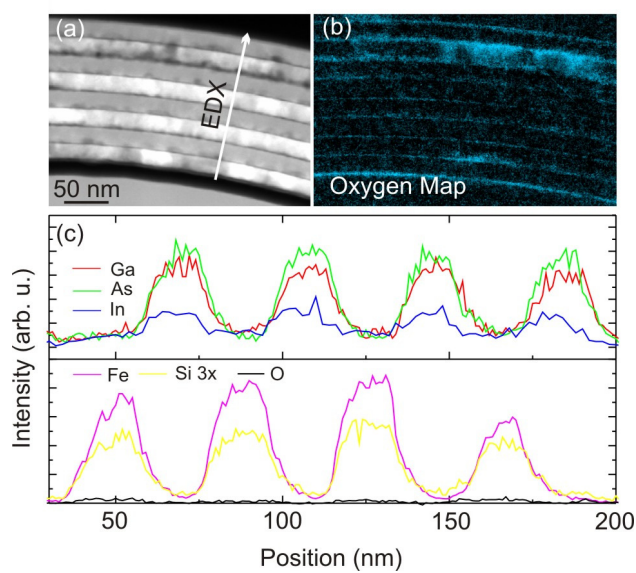

**Figure 1: (color online) (a) HAADF-STEM image of four periods of the radial InGaAs/Fe$_3$Si superlattice. Different materials can be identified by the image contrast. The linescan of the chemical analysis is marked by the white arrow. (b) EFTEM image at the O-K-position illustrating the oxygen distribution in the radial superlattice. (c) EDX linescan across the RSL. The four periods of the superlattice stack can clearly be identified. An oxygen signal is hardly detected.**

Figure 1(a) displays a HAADF-STEM image of four periods of the rolled-up InGaAs/Fe$_3$Si RSL. The two different materials forming the RSL are clearly identified by their different mass-thickness contrast and are tightly bond. From the STEM image the layer thickness is measured to be 19 to 20 nm for both layers in good agreement with the expected layer thickness from growth. The slightly darker InGaAs layer appears more homogeneous than the Fe$_3$Si layer, which exhibits some texture.[4] Surprisingly, no distinguishable interface region is observed, as would be expected from other semiconductor/metal RSLs[13] or semiconductor RSLs.[8-11] To verify the absence of a particular interface region, EFTEM of the O-K-edge as well as EDX were carried out over the layer stack and the results are displayed in Fig 1(b) and 1(c), respectively. The chemical analysis of the RSL (see Fig. 1(c) plot of EDX counts as a function of electron beam position) only shows a clear signal for the elements Ga, In, As, Fe and Si.

These elements oscillate periodically as expected for the RSL structure, easily allowing to discriminate between the two material classes forming the superlattice. The slight drop in the In concentration observed for the semiconductor layer near the InGaAs/Fe$_3$Si interface is attributed to the thermal deoxidation during the Fe$_3$Si deposition. During this process the InGaAs layer temperature exceeds the Indium desorption temperature[16,17] resulting in a Ga rich surface. Additionally, the oxygen content was monitored during EDX analysis. In contrast to other semiconductor/metal hybrid RSLs,[13] no significant oxide-related peak is found at the InGaAs/metal or the metal/InGaAs interface (succeeding layers). Only a slight increase of the oxygen background is noticed in the middle of the Fe peaks. This rise is assigned to an overlap of the Fe and the O signal during EDX and not to a presence of oxygen in the layer. The false color EFTEM image (Fig 1(b)) supports the interpretation revealing only a faint oxide presence at the outer and inner boarders of the whole superlattice as well as in one small region of the Fe$_3$Si RSL. Both inner interfaces (the epitaxial InGaAs/Fe$_3$Si interface and the bonded interface between Fe$_3$Si and InGaAs) show hardly any presence of oxide.

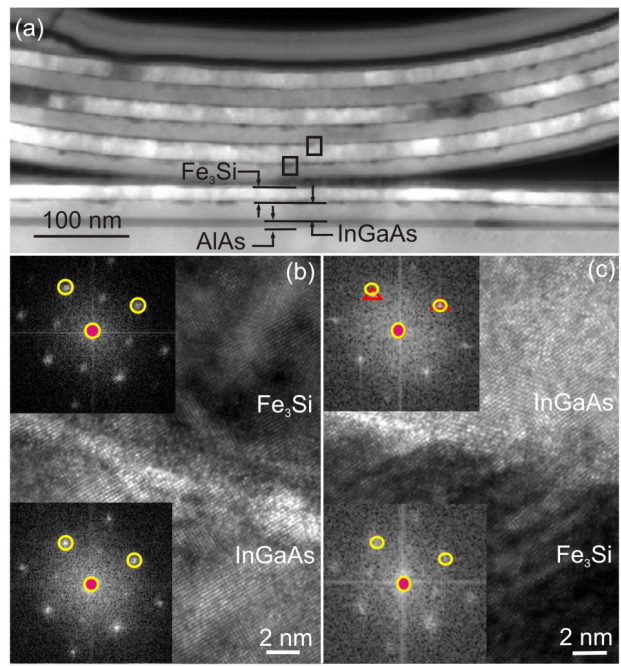

**Figure 2: (color online) (a) TEM image of the InGaAs/Fe$_3$Si radial superlattice as well as the unreleased layer. The positions of the HRTEM images are marked by rectangles. (b) HRTEM of the InGaAs/Fe$_3$Si growth interface as well as the FFT of the lattices of the**

**InGaAs and the Fe$_3$Si layers. Both, the HRTEM image and the FFT indicate a clear pseudomorphic interface of the two layers and a corresponding relation of the lattice. (c) HRTEM image of the Fe$_3$Si and the InGaAs bonding interface. Again the inset gives the FFT of the two lattice images. Note the direct crystalline bonding and the small misalignment indicated by the FFT between the adjacent windings of the radial superlattice.**

To investigate the crystal quality of the two interfaces inside the RSL, high-resolution TEM (HRTEM) was carried out in the lower part of the InGaAs/Fe$_3$Si RSL. A HAADF-STEM image of the investigated area is displayed in Fig. 2(a). Four periods belonging to subsequent turns in the tube as well as the part of the flat initial bilayer of the RSL can be identified. The RSL still lies on the original undetached layer, which is marked together with the unetched AlAs sacrificial layer in the image. HRTEM images of the lower InGaAs/Fe$_3$Si hetero-interface were obtained in the area marked by the lower box in Fig 2(a) and a typical result is shown in Fig. 2(b). The layer interface can be identified as a bright thin line between the lighter InGaAs layer and the darker Fe$_3$Si layer. The lattice fringes running without any disturbance over the interface indicate the pseudomorphic lattice match of both layers. Fast Fourier transformations (FFT) of the InGaAs as well as the Fe$_3$Si part of the HRTEM image are shown in the insets. We marked the central spot and two dominating peaks in the FFT to illustrate the lattice spaces and the orientation of the lattice fringes. The FFT confirms that both layers have the same lattice parameter and crystalline alignment expected for pseudomorphically matched materials, thus manifesting the very good epitaxial relation of our layers. Figure 2(c) shows a HRTEM image of the bonding interface between two succeeding windings of the RSL (area marked by upper box in Fig. 2(a)). Again, the interface is identified by its bright contrast in the middle of the HRTEM image. Surprisingly, undisturbed regions are found in parts of the image were the lattice fringes show no breaks indicating a lattice matched, crystalline bonding of the two layers. To confirm the good lattice alignment of the two succeeding periods of the RSL, FFTs are derived for the Fe$_3$Si and the InGaAs layer and depicted in the two insets of Fig. 2(c). As in Fig. 2(b) we marked the central spot as well as two dominating peaks in the FFT. When compared, only a slight mismatch between the angular alignments of the two FFTs is observed (red triangle and yellow circle), reinforcing the conclusion of a highly lattice matched bonding between the two periods in this area.

To obtain complete crystalline interfaces, the lattice mismatch $\delta$ between succeeding windings has to be small enough to accommodate elastically any strain in the crystalline layers. For the RSL, the lattice mismatch is defined as $\delta=\Delta a/a_{t,o}$, where $\Delta a$ is the lattice difference between the outer tangential lattice parameter $a_{t,o}$ and the inner lattice parameter $a_{t,i}$ of the next winding. Using the linear dependence of $a_t$[18] and assuming that the lattice parameters will not significantly change from winding to winding, $\delta$ depends on the layer thickness $d$ and the radius $R$ (curvature $1/R$) of the rolled-up tube:

$$\delta = \frac{d}{R+d} \qquad (1).$$

Since $R$ can easily be calculated,[19] if the tube reaches the mechanical equilibrium, and $R$ and $d$ are accessible by direct methods, $\delta$ can be determined theoretically as well as from our SEM/TEM data ($\delta$=2.7% for the theoretical growth parameter and tube diameter, $\delta$=3.2% determined from the measured SEM/TEM values). Complementary to this, x-ray diffraction offers direct access to lattice parameters[15,20] providing the possibility to directly measure $\delta$.

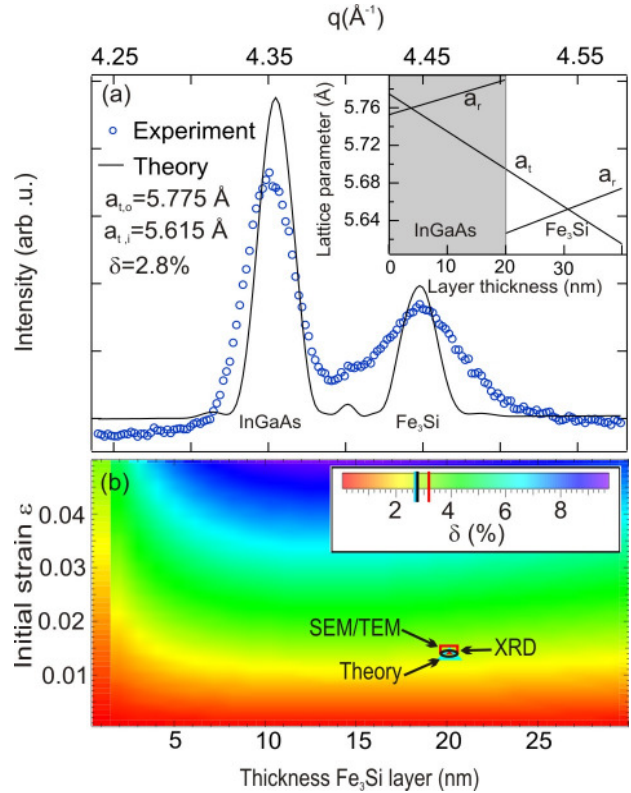

**Figure 3: (color online) (a)** XRD intensity of a single InGaAs/Fe$_3$Si rolled-up microtube (dots). The theoretical solid line is fitted to the experimental data to deduce the lattice parameters of the tube (see inset) **(b)** Color encoded map of $\delta$ as a function of Fe$_3$Si layer thickness and the initial strain in the bilayer. The positions of the theoretical as well as experimental values determined for $\delta$ in the investigated InGaAs/Fe$_3$Si RSL are marked by the red box (SEM/TEM), black circle (XRD) and blue (theory) triangle in the figure.

A measured XRD intensity curve given as a function of the coplanar diffraction vector $q = 4\pi \sin(\theta)/\lambda$ (where $\theta$ is the diffraction angle and $\lambda$ the wavelength) in the vicinity of the (004) reflection of a rolled-up InGaAs/Fe3Si RSL is shown in Fig. 3(a). As expected for rolled-up bilayers,[20] two separate peaks – arising from the two different radial lattice parameters – corresponding to the InGaAs layer ($q=4.35$ Å$^{-1}$) and the Fe$_3$Si layer ($q=4.45$ Å$^{-1}$) are observed. According to Ref. 15, the position and peak shape of the XRD curve is determined by three parameters: (i) the layer thicknesses ($d_1$ and $d_2$), (ii) the tangential and radial lattice parameter distribution through out the tube wall ($a_t$ and $a_r$, respectively) and (iii) the radius $R$. Hereby, $d_1$ and $d_2$ as well as $R$ mainly influence the peak height and width as well as the peak spacing. The peak positions are mainly determined by $a_t$ and $a_r$. Therefore, by fitting $a_r$ and $a_t$ to the peak positions, while taking $d_1$, $d_2$ and $R$ as fixed constrains, $\delta$ can be determined independently from Eq. 1. Furthermore, the quality of the fit using the experimentally measured values for $d_1$, $d_2$ and $R$ provides an indication of the overall crystal quality of our grown InGaAs/Fe$_3$Si bilayer. Using the SEM determined tube diameter (1200 nm) as well as the layer thicknesses determined by TEM (20 nm each), we have evaluated[15] $a_t$ and $a_r$ (inset Fig. 3(a)) as well as the diffraction curve (solid line in Fig. 3(a)) by a least square fit. For the calculation of the lattice parameter profile and the subsequent tube diffraction, we assume that the lattice parameter of Fe$_3$Si is matched in the unrolled layer with the GaAs substrate, within our experimental resolution of ±0.008 Å, for both in-plane and out-of plane directions. This assumption is supported by the observed overlap of GaAs and Fe$_3$Si (004) diffraction peaks in the flat layers (not shown here). The positions of the measured XRD peaks are well described by the theoretical line and using the calculated $a_t$, we determine $\delta$ to 2.8%. The width of the InGaAs peak is similar to the theoretical calculation whereas the Fe$_3$Si measured peak width is broader. We attribute this experimentally observed broadening to a slight in-plane crystalline disorder of the metallic Fe$_3$Si phase. This disorder arises from the lower overall crystal quality of a metallic layer grown on top of the high quality semiconductor crystal template grown by MBE. Nevertheless, a fairly good epitaxial quality is not only indicated by the observation of the Fe$_3$Si peak in tube XRD, but also by the absence of any relaxed Fe$_3$Si peak for the flat bilayers.[21] A clear pseudomorphic relation observed in the HRTEM image of the expitaxial interface is also observed all over the tube (not shown), but the crystal quality of the Fe$_3$Si is hard to determine solely from the images. The slight misalignment of succeeding windings observed in HRTEM – shown in the FFT of the HRTEM images in Fig. 2(c) – breaks the crystalline coherence of the RSL. As discussed in detail in Ref. 15, such full crystal coherence is needed to observe any superlattice superstructures in the RSL-XRD curve.

To compare $\delta$ deduced from TEM, XRD as well as from pure calculations, $\delta$ is plotted in Fig. 3(b) as a function of the Fe$_3$Si layer thickness and the initial strain in the planar layer using Eq. 1 (solutions of Ref. 19 were used to calculate $R$). The initial strain of the planar InGaAs/Fe$_3$Si bilayer is – in a fairly good approximation – defined solely by the lattice mismatch of the InGaAs layer and the GaAs (001) substrate as the grown stoichometric Fe$_3$Si has nearly the lattice constant of GaAs. Furthermore, the three points in Fig. 3(b) mark the $\delta$ found by the different approaches for the investigated RSL. The lower limit (2.7%) is given by using the theoretically calculated $R$ (1400 nm) and total layer thickness $d$ of 40 nm, while the upper limit (3.2%) is obtained from the experimental SEM/TEM values of $R$ and $d$ using Eq. 1. The value obtained by XRD for the local lattice mismatch is in between these two values, denoting a good agreement of all methods. The XRD determined $\delta$ is the most reliable since it accounts for actual curvature of the RSL, with direct access to the local lattice parameter information, as well as the observed layer thickness.

In the literature, stable pseudomorphic growth of $Fe_3Si$ on GaAs (001) has been reported[22,23] for a Si concentration of 10-26% and with a resulting $\delta$ of -0.3 to 2.2% towards its GaAs (001) substrate. If we shift $\delta$ by decreasing the initial Indium content in the InGaAs layer or the $Fe_3Si$ layer thickness into such regime, one could envision large areas of perfect crystalline bonding in the RSL structures allowing for crystalline coherent walls.

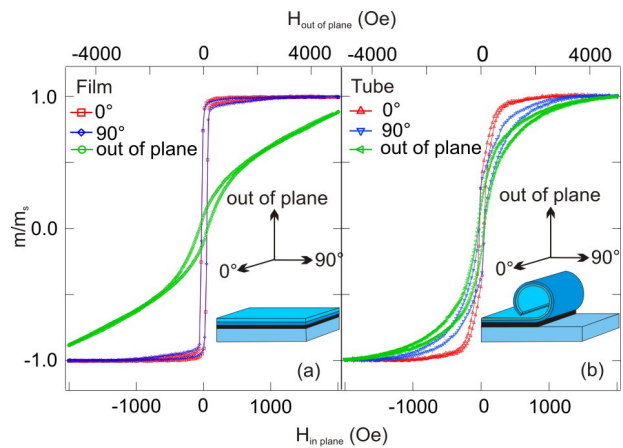

**Figure 4: (color online) (a) Magnetization curves at room temperature of a patterned, unrolled InGaAs/Fe$_3$Si film. (b) Magnetization curves for an area of rolled-up InGaAs/Fe$_3$Si microtubes. Notice the change of the symmetry plane from film to tube due to the cylindrical shape of the rolled-up structures.**

Finally, we measured the magnetization curves of the initial, unetched bilayer and a dense ensemble of rolled-up tubes in a parallel arrangement along the two <010>-directions (0° and

90° to the tube axis; in plane) as well as the [001]-direction (out-of-plane). The measurement geometry is illustrated in the insets of Figs. 4(a) and 4(b). The obtained magnetization curves at room temperature have been normalized to the magnetic saturation moment $m_s$ (planar film 8.6x10$^{-5}$ emu, tubes 6.7x10$^{-5}$ emu) and are plotted in Fig. 4(a) and 4(b), respectively (note the different axes for the in-plane and out-of plane measurements). The planar Fe$_3$Si film exhibits in-plane cubic anisotropy with the easy axis along the <100>-direction in agreement with other reports.[22-24] In contrast to the planar film, the in-plane magnetization curves measured for rolled-up tubes are nonequivalent and no longer exhibit rectangular loops, but instead show smoother behavior, that indicates a harder switching. Furthermore, the shape of the hysteresis curve along the out-of-plane direction is closer to the in-plane measurements and the saturation field is shifted to lower values. These results are attributed to the change in the geometry from a planar film to the cylindrical symmetry of the RSL. This behavior has an analogy with the observed 2D powder behavior of tubes in coplanar x-ray diffraction,[20] where the roll-up of the planar layer makes the tube axis also the unique symmetry axis. In the magnetic case, the tubes keep the properties of the planar film (the easy axis) along this new symmetry axis. All perpendicular directions to this axis should behave (at least qualitatively) similar, as observed in our experiments.

## 4. Conclusions

In conclusion, we have investigated the interfaces, the lattice mismatch and the magnetic properties of a hybrid InGaAs/Fe$_3$Si RSL. EFTEM, EDX and HRTEM as well as the lattice parameter configuration obtained via XRD prove the good epitaxial quality of our initial InGaAs/Fe$_3$Si layers deposited on top of an MBE grown template. Furthermore, areas of perfect bonding are observed by HRTEM between adjacent windings of the RSL. The lattice mismatch $\delta$ between windings is investigated in detail. Our determined lattice mismatch suggests that a full coherent crystalline wall in the magnetic semiconductor/metal hybrid RSL could be feasible by slightly tuning the flat layer parameters. The cylindrical geometry which allows for XRD investigations also changes the behavior and the basic symmetry axis of the magnetization of the hybrid InGaAs/Fe$_3$Si bilayer. Structural as well as the magnetic results let us envision further applications and fundamental effects,[2,3] which are governed by the special cylindrical symmetry of these structures, distinguishing them from their traditional counterparts in planar superlattice geometries.

## Acknowledgements

The authors thank E. Coric, B. Arnold, I. Mönch, S. Sieber, B. Eichler, Y. F. Mei and D. J. Thurmer for experimental help and support. C.D. likes to thank the ID01 staff for their hospitality in Dec. 2007. This work was financially supported by the BMBF (03X5518) and EC NoE Sandie.